\documentclass[12pt]{article}

\makeatletter
\renewcommand{\@seccntformat}[1]{%
	{\csname the#1\endcsname}\ \ }

\renewcommand{\section}{%
         \@startsection{section}{1}{\z@}%
         {-3.5ex plus -1ex minus -.2ex}%
         {2.3ex plus.2ex}%
         {\centering\normalsize\bfseries}}

\renewcommand{\subsection}{\@startsection{subsection}{2}{0pt}%
	{-3.25ex plus -1ex minus -.2ex}%
	{1.5ex plus .2ex}%
	{\centering\normalsize\itshape}}
\makeatother

\newcommand{\Kahler}{K\"a{}hler}

\newcommand{\avg}[1]{\langle #1 \rangle}

\newcommand{\beq}{\begin{equation}}
\newcommand{\eeq}{\end{equation}}

\newcommand{\kahler}{K\"ahler\ }
\newcommand{\nc}{\newcommand}

\newcommand{\Sigmabar}{\bar \Sigma}
\newcommand{\Tbar}{\bar T}
\newcommand{\Hbar}{\bar H}

\newcommand{\GeV}{~\hbox{\rm GeV}}
\newcommand{\mgut}{M_{\rm GUT}}

\begin{document}

\begin{titlepage}

\begin{center}


{\hbox to\hsize {\hfill PUPT-1930}}
{\hbox to\hsize {\hfill UW/PT 00-08}}
{\hbox to\hsize {\hfill UMD-PP-00-79}}
\bigskip


\bigskip

\bigskip

\centerline{\Large\bf The GUT Scale and Superpartner Masses}
\medskip
\centerline{\Large\bf from Anomaly Mediated Supersymmetry Breaking}

\bigskip\bigskip

Z. Chacko$^{\bf a}$, Markus A. Luty$^{\bf b}$, Eduardo
Pont\'on$^{\bf b}$,
Yael~Shadmi$^{\bf c}$, and Yuri~Shirman$^{\bf c}$
\bigskip

$^{\bf a}${\small \it Department of Physics, Box 351560\\
University of Washington\\
Seattle, Washington 98195\\
{\rm zchacko@fermi.phys.washington.edu}}
\smallskip

$^{\bf b}${\small \it Department of Physics\\
University of Maryland\\
College Park, Maryland 20742\\
{\rm mluty@physics.umd.edu, eponton@wam.umd.edu}}

\smallskip

$^{\bf c}${\small \it Department of Physics\\
Princeton University\\
Princeton, New Jersey 08544\\
{\rm yshadmi@feynman.princeton.edu, yuri@feynman.princeton.edu}}

\bigskip
{\bf Abstract}
\medskip\noindent
\begin{minipage}[t]{5.3in}%
We consider models of anomaly-mediated supersymmetry breaking
(AMSB) in which the grand unification (GUT) scale is determined by the
vacuum expectation value of a chiral superfield.
If the anomaly-mediated contributions to the potential are
balanced by gravitational-strength interactions, we find
a model-independent prediction for the GUT scale of order
$M_{\rm Planck} / (16\pi^2)$.
The GUT threshold also affects superpartner masses,
and can easily
give rise to realistic predictions if the GUT
gauge group is asymptotically free.
We give an explicit example of a model with these features,
in which the doublet-triplet splitting problem is solved.
The resulting superpartner spectrum is very different from
that of previously considered AMSB models, with gaugino masses
typically unifying at the GUT scale.
\end{minipage}

\end{center}

\end{titlepage}

\renewcommand{\thepage}{\arabic{page}}
\setcounter{page}{1}

\section{Introduction}
\label{introduction}
In supersymmetric extensions of the Standard Model,
supersymmetry (SUSY) is broken in some separate sector of the theory
and communicated to visible sector fields through gauge
or gravitational interactions.
Gravitational effects typically lead to superpartner masses from
contact terms suppressed by powers of the Planck scale, giving rise
to superpartner masses of order $m_{3/2}$.
If these contact terms are absent, superpartner masses of order
$m_{3/2} / (16\pi^2)$ are
still
generated due to the superconformal anomaly
\cite{rs,glmr}.
This is called anomaly-mediated SUSY breaking (AMSB), and it
naturally dominates
if SUSY is broken on a separate brane~\cite{rs,ls}.
In AMSB, all soft masses are related to beta functions and anomalous
dimensions,
and are therefore completely determined by
the quantum numbers of the relevant field up to
the overall scale.
In particular,
squark masses other than the stop are determined to high
accuracy by the gauge couplings, and therefore there are no dangerous
flavor-changing neutral currents (FCNC's).

The minimal version of AMSB is obtained by simply coupling the
minimal supersymmetric standard model (MSSM) to AMSB.
However, this gives rise to negative slepton mass-squared terms
at the weak scale due to the signs of the $SU(2)_W$ and $U(1)_Y$
beta functions.
Modifying
the superpartner spectrum
is nontrivial because the AMSB predictions are
largely
insensitive to heavy supersymmetric thresholds.
Several solutions to this problem have been
suggested~\cite{clmp,kss,rp,jj,rparity}.
One is to change the effective theory at the
weak scale to an extension of the MSSM~\cite{clmp}.
Another possibility relies on the fact that any heavy threshold
affects the soft terms beyond leading order in the SUSY breaking.
If the SUSY-breaking splittings of a heavy multiplet are large,
there can be large threshold corrections to visible sector sparticle
masses~\cite{kss}.
Finally,
Ref.~\cite{rp} pointed out that
if a heavy threshold is determined by a modulus field whose potential
arises from SUSY breaking, the predictions for visible sparticle
masses are changed.
This latter mechanism is particularly attractive, and is the one we
will employ in the present paper.

An obvious candidate for the heavy threshold is the grand unification
(GUT) scale.
In fact, we will show that there is a simple and robust mechanism in
which AMSB gives rise to a VEV for a modulus field of order
$M_{\rm Planck} / (16\pi^2)$, precisely the right size of the GUT scale.
(The mechanism is very similar to a mechanism first discussed in
the context of gauge mediated models~\cite{clp}. Other mechanisms
for dynamically generating the GUT scale were considered
in \cite{othergut}.)
In addition, this threshold
can lead to a realistic superpartner spectrum.
We present an explicit model that incorporates these features, while
also explaining doublet-triplet splitting.

The phenomenology of this class of models is very interesting.
The soft SUSY breaking masses can be obtained by starting with the
AMSB soft masses at the GUT scale, and then running them down to the
weak scale without large threshold corrections at the GUT scale.%
\footnote{If the GUT threshold were supersymmetric, the threshold corrections
would be large, resulting in minimal AMSB predictions for the low-energy
theory.}
Therefore, the gaugino masses generally unify at the GUT scale as in
traditional
hidden sector models or gauge-mediated models.

One constraint on this class of models is that extra structure is
required to avoid FCNC's.
The problem is that dimension-4 \Kahler\ terms can mix the modulus
field with visible sector fields, giving rise to visible sector scalar
masses of order $\avg{X}^2 m_{3/2}^2 / M_{\rm Planck}^2$, where
$\avg{X}$ is the modulus VEV.
For $\avg{X} \sim \mgut \sim 10^{16}\GeV$, this is as large
as the AMSB contribution, and there is no reason for this
contribution to conserve flavor.
This constraint applies to any model with a large modulus VEV that
is determined by SUSY breaking.
We will show that these terms can be naturally suppressed by assuming
that the visible sector gauge and Higgs fields live on a `thick brane',
with the hidden sector and the visible sector matter fields localized
at different positions within the brane.%
\footnote{The fact that `thick branes' can modify naturalness in an
interesting way was pointed out in Ref.~\cite{as}.}
Alternatively, one can simply assume that the unwanted dimension-4
terms are suppressed;
this does not appear as unnatural as the fine-tuning of conventional
hidden-sector models, since we assume that \emph{all} \Kahler\ terms
of a certain structure are small.

Because of the constraint above, we also briefly consider mechanisms
that give lower thresholds.
These can be motivated by GUT's with intermediate scales.
The results depend on the details of the models, but we conclude that
there is a large class of interesting models.

This paper is organized as follows.
In Section~\ref{amsb},
we review some basic facts about AMSB and the non-decoupling
of heavy thersholds that are determined by moduli.
In Section~\ref{gutscale}, we explain our mechanism of obtaining the GUT scale
dynamically from the Planck scale.
We also discuss several alternatives for generating a non-decoupling threshold.
We construct a specific model in Section~\ref{model}.
Several possible mechanisms for generating a $\mu$ term are
described in Section~\ref{mu}.
We discuss phenomenology in Section~\ref{phenom}.
Our conclusions are in Section~\ref{concl}.

\section{AMSB and Decoupling}
\label{amsb}

If SUSY is broken in a sector that communicates with the visible
sector only via gravity, and there are no contact interactions
between the hidden and visible sector, then SUSY breaking will
be communicated by anomaly mediation.
In the visible sector, all SUSY breaking effects enter through the
supergravity chiral compensator field
\beq
\phi = 1 + \theta^2 F_\phi,
\eeq
where 
$F_\phi \sim m_{3/2}$.
The couplings of $\phi$ are restricted by the fact that $\phi$ is
chiral, and by
a spurion scale symmetry under which $\phi$ has mass dimension $+1$.
Therefore, $\phi$ only appears in terms with dimensionful couplings.
For example, a chiral mass term is covariantized by the replacement
$M \rightarrow M \phi$.

If the visible sector does not contain dimensionful couplings, there
is no SUSY breaking at tree level.
However, in a supersymmetric regulator, the cutoff is a dimensionful coupling
that must be covariantized. 
This means that there are soft masses at loop level.
Writing the Lagrangian for the visible sector
\beq
{\cal L}_{\rm vis} = \int\! d^4\theta\, Z(\mu/\Lambda) Q^\dagger e^V Q
+ \left( \int\! d^2\theta \left[ S(\mu/\Lambda) W^\alpha W_\alpha
+ \lambda Q^3 \right]
+ \hbox{\rm h.c.} \right),
\eeq
we replace
\beq
Z\!\left( \frac{\mu}{\Lambda} \right)
\to Z\!\left( \frac{\mu}{\Lambda (\phi^\dagger \phi)^{1/2}} \right),
\qquad
S\!\left( \frac{\mu}{\Lambda} \right) \to S\!\left(
\frac{\mu}{\Lambda \phi} \right).
\eeq
Expanding in $\theta$, one finds the AMSB soft
masses%
\footnote{This is analogous to the method of
Giudice and Rattazzi~\cite{gr} for extracting soft masses in models
of gauge mediated supersymmetry breaking.}
\begin{eqnarray}
\label{soft}
m_0^2(\mu) \!\!&=&\!\! -\frac{1}{4}\, \frac{\partial
\gamma(\mu)}{\partial\ln\mu}\,
\vert F_\phi \vert^2\ , \cr
m_{1/2}(\mu) \!\!&=&\!\! \frac{\beta(\mu)}{g} F_\phi\ , \cr
A(\mu) \!\!&=&\!\! -{1\over 2}\gamma(\mu) F_\phi \ .
\end{eqnarray}
Here $\beta$ is the gauge beta function and
$\gamma = \partial\ln Z / \partial \ln \mu$ is the anomalous dimension.

It is not hard to see why a supersymmetric threshold does not affect
the soft scalar masses.
For example, the gaugino masses at one loop are parameterized by the
chiral superfield
\beq
S = \frac{1}{2 g^2} + \frac{m_{\lambda}}{g^2} \theta^2\ .
\eeq
The effective gauge coupling below a SUSY threshold $M$ is given by
\beq
S_{\rm eff}(\mu) = S_0 + \frac{b}{16\pi^2} \ln \frac{M}{\Lambda}
+ \frac{b_{\rm eff}}{16\pi^2} \ln \frac{\mu}{M}\ ,
\eeq
where $b$ and $b_{\rm eff}$ are the beta function coefficients in
the theory above and below the scale $M$, respectively.
To covariantize, both $M$ and $\Lambda$ are rescaled by $\phi$:
\beq
S_{\rm eff}(\mu) \to S_0 + \frac{b}{16\pi^2} \ln \frac{M}{\Lambda}
+ \frac{b_{\rm eff}}{16\pi^2} \ln \frac{\mu}{M \phi}\ .
\eeq
We see that the soft masses depend only on the beta function in
the effective theory below $M$.
This result generalizes to all couplings to all orders in perturbation
theory: the effective coupling is a function of the ratios
$\mu / M \to \mu / (M\phi)$ and $M / \Lambda \to M / \Lambda$.
The threshold $M$ acts as the cutoff for the effective theory,
and the soft masses are insensitive to the presence of the heavy
threshold.%
\footnote{This only holds to leading order in the SUSY
breaking. The soft terms typically receive corrections of order
$F^4/M^2$ and higher due to the heavy threshold.
These corrections are of course only relevant for
small $M$.\cite{kss}}

It is worth reviewing how these results arise in terms of component
diagrams.
An explicit SUSY mass term $M$ will give rise to a SUSY breaking
$B$-type mass-squared term $M F_\phi$.
Loop threshold corrections at the scale $M$ therefore give SUSY breaking masses
of order $F_\phi / (16\pi^2)$,
precisely the size of the AMSB contributions.
These corrections ensure that the low-energy SUSY breaking masses
coincide with the AMSB predictions appropriate to the effective
theory below the scale $M$.

The results are completely different if the threshold is due to the
VEV of a light field whose potential arises from SUSY breaking~\cite{rp}.
The models we consider have a field $X$ with superpotential couplings of
the form
\beq
\Delta W = \lambda X T_1 T_2.
\eeq
In the SUSY limit, there is a flat direction with $X \ne 0$
along which the
fields $T_{1,2}$ get masses $\lambda X$.
Integrating out $T_{1,2}$ at the scale $\langle X\rangle$
we can read the gaugino mass off the low-energy gauge coupling function
\beq
S_{\rm eff}(\mu) \to S_0 + \frac{b}{16\pi^2} \ln \frac{\lambda X}{\Lambda\phi}
+ \frac{b_{\rm eff}}{16\pi^2} \ln \frac{\mu}{\lambda X}\ .
\eeq
Since $X$ is a field, it is not rescaled by $\phi$,
so that $\langle F_X\rangle\neq \langle X\rangle \langle F_\phi\rangle$,
and the AMSB prediction is modified.
Models of this type are considered in Ref.~\cite{rp}.

As we explain in the next Section,
in the models we construct
$\avg{F_X} / \avg{X} \ll F_\phi$.
Then, the soft masses at the weak scale can be obtained by
calculating the AMSB soft masses in the full theory
at the scale $\langle X\rangle$
and running down to the weak scale.
This can be seen simply by considering the component calculation of the soft
masses.
The fields that are integrated out at the GUT scale have SUSY masses of
order $\avg{X}$, and SUSY breaking mass-squared terms of order $\avg{F_X}$.
Because $\avg{F_X} / \avg{X} \ll F_\phi$, integrating out these fields
does not give large loop matching corrections, and tree level matching
with 1-loop running gives a good approximation to the low-energy masses.

\section{The GUT Modulus}
\label{gutscale}
We now describe the mechanism for generating the GUT scale.
We consider GUT models with a flat direction
along which the GUT gauge group is spontaneously broken to the
Standard-Model gauge group.
We also require that AMSB give a negative mass-squared to this flat
direction; we will discuss this point further below.
If there are no other contributions to the potential for the flat
direction, the
theory will run away, presumably giving a noninteracting theory.
However, the potential can be stabilized by \Kahler\ terms of the form
\beq\label{eq:stabkahler}
\Delta{\cal L} = \int\! d^4\theta\,
\frac{c}{M_{\rm Planck}^2 \phi^\dagger \phi} (\Sigma^\dagger \Sigma)^2\ ,
\eeq
where $\Sigma$ is a field with a nonzero VEV along the flat direction.
This gives a contribution to the potential for $\Sigma$
\beq
\Delta V(\Sigma) = -\frac{c \vert F_\phi \vert^2}{M_{\rm Planck}^2}
(\Sigma^\dagger \Sigma)^2\ .
\eeq
For $c < 0$ this can stabilize the VEV at
\beq
\avg{\Sigma} \sim \left(
\frac{M_{\rm Planck}^2 m^2_\Sigma}{c \vert F_\phi \vert^2} \right)^{1/2}
\sim \frac{M_{\rm Planck}}{16\pi^2}\ ,
\eeq
where $m_{\Sigma}^2 \sim \vert F_\phi \vert^2 / (16\pi^2)^2$ is the
AMSB soft mass of $\Sigma$.
Note that this automatically generates a VEV at the GUT scale,
assuming only that all couplings are order 1.
The correct value of the GUT scale is
generated from the Planck scale purely
by supergravity effects!

AMSB dominates only if there are no large contact terms between the
hidden and observable sector.
This is natural if the hidden and observable sector are localized on
different branes, and separated by a distance that is larger than the
fundamental Planck scale.
But if there are large extra dimensions, the size of higher-dimension
operators such as Eq.~(\ref{eq:stabkahler}) is set by the fundamental
Planck scale, which is smaller than the 4-dimensional Planck scale.
For one extra dimension, this relation is
\beq
M_{\rm Planck}^2 = 2\pi r M_5^3\ ,
\eeq
where $r$ is the radius of the extra dimension.
The unwanted direct contact terms have the form
\beq
\Delta{\cal L} = \int\! d^4\theta\, \frac{a_{jk}}{M_5^2}
N^\dagger N Q^\dagger_j Q^{\vphantom\dagger}_k\ ,
\eeq
where $N$ is the hidden sector field that breaks SUSY via $\avg{F_N} \ne 0$,
and $a \sim e^{-M_5 r}$.
It is easy to see that for $r \sim 10 / M_5$ we can suppress the unwanted
operators and still obtain a reasonable value of the GUT scale via the
mechanism above.

We now turn to the question of how the threshold at $\avg{\Sigma}$ affects
the SUSY breaking masses.
The relevant
quantity is $\avg{F_\Sigma} / \avg{\Sigma}$, since this
determines the size
of the SUSY breaking corrections from loops
involving GUT scale fields.
The effective theory below the scale $\avg{\Sigma}$ contains
a light Standard Model singlet field that parametrizes
the motion along the flat direction. We label this field by $X$.
(In the fundamental theory, the flat direction typically corresponds to
a combination of fields, some of which have GUT charges.)
In the effective theory, the potential for $X$ is
given by
\beq\label{eq:Phipot}
\Delta{\cal L}_{\rm eff} = \int\! d^4\theta \left[
Z_X\!\left( \mu = (X^\dagger X)^{1/2}\right)
X^\dagger X
+ \frac{c}{M_{\rm Planck}^2 \phi^\dagger \phi} (X^\dagger X)^2
\right]\ .
\eeq
The wavefunction factor for $X$ is evaluated at the scale
$\mu = (X^\dagger X)^{1/2}$ because $X$ has only irrelevant interactions
below the scale $\avg{\Sigma}$ where the GUT fields are integrated out.
$\avg{F_X}$ is determined by expanding the first term
\beq
\Delta{\cal L}_{\rm eff} \sim F_X^\dagger F_X
+ \left( F_X^\dagger \frac{X\, F_\phi}{16\pi^2} + \hbox{\rm h.c.} \right)\ ,
\eeq
which gives
\beq
\frac{\avg{F_X}}{\avg{X}} \sim \frac{F_\phi}{16\pi^2}\ .
\eeq
Since $\avg{F_X} \ll \avg{X}F_\phi$, we can neglect the contributions
from $X$ as a messenger field compared to the contributions from AMSB.
As discussed earlier, the soft masses can therefore be computed by running
the AMSB masses at the GUT scale down to the weak scale.%
\footnote{The renormalization group running is also suppressed by a factor
of $1/(16\pi^2)$, but is enhanced by large logs.
There are no large logs in the $X$ messenger contribution, since the
parameters in Eq.~(\ref{eq:Phipot}) are not renormalized below the GUT
scale.}

The gauge contribution to the one-loop running of the
scalar masses squared is always positive.
However, the positive running contribution to the right-handed slepton
masses is very small, so we are led to require that the scalar masses
are positive at the GUT scale.
The AMSB contribution to scalar masses has the schematic form
\beq\label{eq:mform}
m^2 \sim \left( \frac{1}{16\pi^2} \right)^2
\left[ +y^2 (y^2 - g^2) \pm g^4 \right]\, F_\phi^2,
\eeq
where $y$ is a Yukawa coupling and $g$ is a gauge coupling.
The last term is positive if the gauge group is asymptotically free,
and negative if it is infrared free.
For the first two generations, the Yukawa couplings are small.%
\footnote{One could attempt to build models in which the leptons have
large Yukawa couplings with exotic fields above the GUT scale.
We will not pursue this possibility here.}
We therefore require the slepton fields to be charged under an
asymptotically free gauge group.
The only possibilities are a gauged non-abelian horizontal symmetry, or the GUT
group itself.
However, the number of GUT fields is so large that a gauged horizontal symmetry
will not be asymptotically free.
We are therefore led to the requirement that the GUT group be asymptotically
free.
It is nontrivial to satisfy this requirement in a realistic GUT, but we will
give an example in the next section that shows that it is possible.

If the GUT group is asymptotically free, then the $g^4$
contribution to the mass-squared of the GUT modulus is also
positive, while our mechanism requires a negative value.
We can obtain a negative mass-squared for the flat direction
either from the $-y^2 g^2$ term in Eq.~(\ref{eq:mform}), or
by charging
some of the fields along the flat direction
under an infrared-free gauge group (such as a $U(1)$ factor).
Both of these effects are present in the model of Section~\ref{model},
and give a negative mass-squared for the flat direction.

Since it is difficult to construct a realistic asymptotically free
$SU(5)$ GUT model, larger groups are typically needed.
Thus, the Standard Model fields
will generically be charged under some broken diagonal generators of the
GUT group. As discussed in~\cite{rp} this can give rise to
additional $D$-term contributions to the soft masses.
These contributions can have important phenomenological
consequences, but they are highly model dependent.
We note that it may be possible to construct a model in which
$D$-term contributions to slepton masses squared are sufficiently
positive, so that the GUT group does not have to be asymptotically
free. For this to work, all sleptons must have same-sign charges
under the appropriate $U(1)$. We do  not explore this possibilty here.

An important constraint on this scenario is that we must forbid
couplings of the form
\beq\label{eq:badop}
\Delta{\cal L} = \int\! d^4\theta
\frac{c_{jk}}{M_{\rm Planck}^2 \phi^\dagger \phi}
(\Sigma^\dagger \Sigma) (Q^\dagger_j Q^{\vphantom\dagger}_k)\ ,
\eeq
where the $Q$'s are MSSM matter fields.
These terms cannot be forbidden by any symmetries, and there is no reason
that they should conserve flavor.
They give rise to contributions to scalar masses of order
\beq
\Delta m^2_{jk} \sim \frac{c_{jk} |F_\phi|^2}{M_{\rm Planck}^2}
\avg{\Sigma}^\dagger \avg{\Sigma}
\sim c_{jk} \left( \frac{1}{16\pi^2} \right)^2 |F_\phi|^2\ ,
\eeq
where we have used $\avg{\Sigma} \sim M_{\rm Planck} / (16\pi^2)$.
These contributions are precisely the size of the AMSB
contributions, ruining the natural absence of FCNC's.
This fine-tuning
is in a sense less severe than in traditional hidden-sector models,
because we require \emph{all} operators of the form Eq.~(\ref{eq:badop})
to be small,
and not just a special subset of them
(the flavor non-diagonal ones).

We can suppress the dangerous contributions without fine tuning
by localizing the GUT modulus fields
$\Sigma$ and the
visible sector fields $Q$ on different branes.
Locality then guarantees the absence of terms such as Eq.~(\ref{eq:badop}).
Gauge fields must couple to both $\Sigma$ and $Q$, and must therefore
live in the bulk.
Doublet-triplet splitting typically
requires that the Standard Model
Higgs fields couple to $\Sigma$ (the field that breaks the GUT group) as well
as to $Q$.
The Standard Model Higgs fields must therefore also live in the bulk.
The  simplest version of these ideas therefore allows the Standard
Model gauge and Higgs fields
to couple directly to the hidden sector where SUSY is broken.
This gives soft terms
that are large compared to the AMSB
contributions.%
\footnote{Putting the gauginos and Higgs fields in the bulk and allowing
direct couplings to the hidden sector results in `gaugino mediated
SUSY breaking'~\cite{gmsb}, which gives a natural and interesting scenario.
However, we are
interested in exploring the possibility that AMSB dominates.}

The way out is to assume that the Standard Model gauge and Higgs fields
are localized on a `thick brane', with the matter fields $Q$ and GUT
Higgs fields $\Sigma$ localized at different positions within the
thick brane.
The hidden sector is assumed to be localized on a brane outside the
thick brane.
Now (approximate) locality  suppresses
all contact terms between the hidden
and visible sector, while also suppressing the undesirable couplings
among the visible sector fields.
This structure arises in
simple models with scalar domain walls~\cite{as}.%
\footnote{For further discussion of SUSY models involving thick
branes see for example Ref.~\cite{kt}.}

We close this section by outlining a few alternatives for
dynamically generating a non-decoupling threshold.
The first is to generate $m^2_\Sigma < 0$ as above, but to stabilize
$\Sigma$ with superpotential terms of the form
$$\Delta W \sim {\Sigma^n\over M_{\rm Planck}^{n-3}}\ .$$
(Here we loosely treat $\Sigma$ as a gauge-singlet for simplicity.
In concrete models $\Sigma$ is typically replaced by a pair of fields
which parameterize a $D$-flat direction.)
Such a superpotential term leads to a scalar potential contribution
of the form
\beq\label{wterms}
\Delta V \sim {\Sigma^{2(n-1)}\over M_{\rm Planck}^{2(n-3)}} +
F_\Sigma {\Sigma^n\over  M_{\rm Planck}^{n-3}}\ .
\eeq+
The second term in~(\ref{wterms}) is subdominant,
and the first term would stabilize the VEV at
$$\langle\Sigma\rangle \sim 10^{-{15 / (n-2)}}
\,M_{\rm Planck}\ .$$
For example, for $n = 4$ we obtain $\avg{\Sigma} \sim 10^{11}\GeV$.
Such superpotential terms may therefore be used to generate
a threshold in models of intermediate scale unification.
We also note that in order to rely on our basic mechanism of using
\kahler potential terms to generate the GUT scale, superpotential
terms of the form~(\ref{wterms}) should be forbidden by some symmetry
for $n\sim 9$ or smaller,
since otherwise they would lead to a VEV smaller
than $\mgut\sim 10^{16}\GeV$.

Finally, another way to generate a non-decoupling threshold is
to use a strongly-coupled theory in which a runaway potential is generated
dynamically, driving some field to infinity.
This field may be stabilized by a positive AMSB scalar mass-squared.
In this case,
the resulting VEV is
related to the SUSY breaking scale as well as
the strong coupling scale of the theory.

\section{The `5--3--1' Model}
\label{model}
We now construct a simple model realizing the ideas discussed in
the previous section.
The most important model building constraint arises from the requirement that
the Standard Model fields are charged under an asymptotically free gauge group
above the GUT scale.
This requirement would be satisfied, for example, in the simple $SU(5)$ model
with only a GUT modulus in addition to the MSSM fields.
However, in any GUT model with an $SU(5)$ factor one needs to generate a
GUT-scale mass for the Higgs triplets.
This requires introducing a number of additional fields, and generically leads
to loss of asymptotic freedom.
A simple solution to the doublet-triplet splitting problem that maintains
asymptotic freedom can be obtained by giving up unification in the strictest
sense.
We follow Ref.~\cite{missing} and consider a model with an
$SU(5)\times SU(3)\times U(1)$ gauge group that is spontaneously
broken to a `diagonal' standard-model subgroup.
The predictions of unification are recovered if the gauge couplings of the
$SU(3)$ and $U(1)$ factors are sufficiently larger than the $SU(5)$ gauge
coupling.

%
\begin{table} \begin{center}
\label{table1}
\begin{tabular}{c|c|c|c} 
$\ $ & $SU(5)$ &  $SU(3)$ & $U(1)$ \\
\hline
$\Sigma$ & 5& $\bar 3$ & 1 \\
$\Sigmabar$ & $\bar 5$& 3 & $-1$ \\
$\Delta$ & 1 & 8 & 0 \\
$T$ & 1& 3 & $-1$ \\
$\Tbar$ & 1& $\bar 3$ & 1 \\
\end{tabular}
\end{center}
\begin{caption}
{Field content of the `5--3--1' model.
The standard-model matter fields are charged under $SU(5)$.}
\end{caption}
\end{table}

The MSSM fields are charged only under the $SU(5)$ factor.
In addition, the model contains the fields of Table~1.
Note that the $SU(5)$ factor is asymptotically free, so AMSB masses for
sleptons and squarks are all positive at the GUT scale.
The superpotential is
\beq
\label{superpot}
W = \lambda_1 \Sigmabar \Delta \Sigma + \lambda_2 \Sigmabar H
\Tbar + \lambda_3 \Sigma \Hbar T \, .
\eeq
This model has a one-parameter flat direction with
\beq
\avg{\Sigma}, \avg{\bar{\Sigma}}
\propto \pmatrix{1 & 0 & 0 \cr 0 & 1 & 0 \cr  0 &0& 1 \cr 0 &0 & 0 \cr
0 &0  & 0\cr}\ ,
\eeq
and all other fields vanishing.
This breaks $SU(5) \times SU(3) \times U(1)$ down to the Standard Model gauge
group, and this flat direction therefore acts as the GUT modulus in this
model.
The role of the other fields is as follows.
$H$ and $\Hbar$ transform as a $5$ and a $\bar 5$ of
$SU(5)$ and contain the Standard Model Higgs doublets.
The fields $T$ and $\bar{T}$ are $SU(3)$ triplets that
get masses with the triplet components of $H$ and $\Hbar$
along the flat direction (the `missing partner' mechanism).
The field $\Delta$ is required to give masses to unwanted
light fields.
In the absence of the $\Delta$ term, there are unwanted Nambu-Goldstone
modes due to accidental global symmetries of the superpotential.
With the couplings above, the only light
fields are the Standard Model fields and a single flat direction that
can be parameterized by $\mathop{\hbox{\rm tr}}( \Sigma \Sigmabar)$.
(For $\Sigma = 0$ there are also flat directions with $\Delta \ne 0$,
but these do not affect the properties of vacua with $\Sigma \ne 0$.)

Unification to $5\%$ accuracy requires the $U(1)$ gauge coupling
to satisfy $g'_1(\mgut) \ge 0.8$.
(We normalize $g'_1$ so that $D_\mu = \partial_\mu - i g'_1 X$, where
$X$ is the $U(1)$ charge defined in Table 1.)
For this value of $g'_1$, the $U(1)$ Landau pole is at $30 \mgut$.
This is uncomfortable, but plausibly large enough to trust perturbation
theory at the GUT scale.

We now consider the effect of SUSY breaking on the GUT modulus flat direction.
The $g_5^4$ contribution to the mass-squared is positive because
$SU(5)$ is asymptotically free, but there are several negative
contributions that can offset this.
First, note that the $g_1^4$ contribution is negative, and can
naturally be larger than the $g_5^4$ contribution because $g_1$ is
required to be large for unification.
(Even at the lower limit of $g_1$, the contributions are comparable.)
The $SU(3)$ factor can also give a negative contribution due to a
fortunate accident of this model:
the one-loop beta function for $SU(3)$ vanishes, so there is no
$g_3^4$ term, and the negative $y^2 g_3^2$ term can naturally dominate
the positive $y^4$ term if $g_3$ is larger than $y$.
We conclude that one can obtain a negative mass-squared for the flat
direction for a wide range of parameters.

The potential for the flat direction
can be stabilized by the \kahler terms
\beq\label{kahler}
\Delta{\cal L} = \int\! d^4\theta\,
\frac{c}{M_{\rm Planck}^2 \phi^\dagger \phi}
(\Sigma^\dagger\Sigma)^2
+ \frac{\bar c}{M_{\rm Planck}^2 \phi^\dagger \phi}
(\Sigmabar^\dagger\Sigmabar)^2 \ ,
\eeq
with $c, {\bar c} \sim 1$.
As discussed in Section~\ref{gutscale},
for $c, {\bar c} < 0$ this gives a stable minimum at
\beq
\avg{X} \sim \frac{1}{16\pi^2} M_{\rm Planck}\ .
\eeq

As discussed in the previous section,  superpotential
terms of the form $(\Sigmabar \Sigma)^n/M_{\rm Planck}^{2n-3}$ must be
forbidden for $n<5$. We note that one can impose an $R$ symmetry
to forbid such operators. Such a symmetry can only be made
anomaly free if the Standard Model fields transform
non-trivially under the symmetry and some of the Standard
Model Yukawa couplings are generated after spontaneous
breakdown of the $R$ symmetry. This is plausible since the
Standard Model Yukawas are small. One could also forbid
dangerous operators by imposing discrete symmetries.

So far we have neglected $D$-term contributions.
However, there are various effects that can give
$\avg{\Sigma} \ne \avg{\Sigmabar}$.
First, note that there is a nonzero AMSB contribution to
$m^2_\Sigma - m^2_{\Sigmabar}$ due to the top-quark Yukawa coupling,
and also if $\lambda_2 \ne \lambda_3$:
\beq
m_\Sigma^2 - m_{\Sigmabar}^2 \sim g_1^2 (\lambda_2^2 - \lambda_3^2)
+ \lambda_2^2 y_t^2.
\eeq
The large $g_1^4$ contributions cancel in the difference,
so it is not unnatural to have
$m_\Sigma^2 - m_{\Sigmabar}^2 \ll m^2_\Sigma + m^2_{\Sigmabar}$.
Another contribution to the difference between $\avg{\Sigma}$
and $\avg{\Sigmabar}$ arises from an asymmetry in the \kahler potential
couplings $c \ne \bar{c}$.

The only non-vanishing $D$ term in our model is associated with the
broken $U(1)$ corresponding to the linear combination of the original
$U(1)$ of the model and the hypercharge subgroup of $SU(5)$.%
\footnote{$D$-terms for other broken diagonal
generators vanish due to the tracelessness of the group
generators and the fact that the modulus VEV is proportional
to the unit matrix under the corresponding subgroup. For example,
for the $SU(3)$ generator $T_8=\mathop{\rm diag}(1,1,-2)$ one has
$\Sigma^\dagger T_8 \Sigma=0$, independently of the value of $\Sigmabar$.}
Thus,  MSSM scalars receive
additional contributions to their soft masses-squared proportional
to their hypercharge.
One should therefore check that these contributions do not
drive the scalar masses negative.
The largest dangerous contribution is the one associated
with either the left-handed or the right-handed sleptons.
It is given by
\beq
\label{dtermcontribution}
m_D^2= Y_\ell \, {g_5^2 \over g_1^2 + g_5^2}
\left[ m^2_\Sigma - m^2_{\Sigmabar} +
\frac{c-\bar c}{c+\bar c} (m^2_\Sigma + m^2_{\Sigmabar}) \right]\,,
\eeq
where $Y_\ell$ is the slepton hypercharge.
If we require unification with
$5\%$ accuracy, we find
\beq
{m_D^2\over m_{\rm AMSB}^2} =
\left(\frac{m^2_\Sigma-m^2_{\Sigmabar}}{m^2_\Sigma + m^2_{\Sigmabar}}+
\frac{c-\bar c}{c+\bar c}
\right) k \,,
\eeq
where $m_{\rm AMSB}^2$ is the usual AMSB mass-squared
and $k \simeq 1$
for both right-handed and left-handed sleptons.
We therefore conclude that in a large
region of parameter space, our model predicts positive slepton
masses at the GUT scale.

\section{The $\mu$ Problem}
\label{mu}
No supersymmetric model is complete without a solution to the `$\mu$ problem.'
In this Section, we briefly consider several approaches in the present
class of models.

The first is the mechanism proposed in Ref.~\cite{rp}.
In this mechanism, one posits a singlet $S$ that gets its mass from the
modulus $X$.
If this singlet has superpotential couplings of the form
\beq
\Delta W =  \lambda S H_u H_d
\eeq
this will generate a $\mu$ term and a $B \mu$ term of the correct size.
As discussed in Ref.~\cite{rp}, there is no danger of generating a large
$B$ term because there is no $F$-type VEV larger than $F_\phi / (16\pi^2)$.

This mechanism can be adapted to the present class of models.
In order for the GUT modulus to give a GUT-scale mass to a singlet, we
add the superpotential terms
\beq
\Delta W \sim N (\bar{\Sigma} \Sigma - Y^2) + Y S^2 + S \bar{H} H + S^3 \ ,
\eeq
where $N, Y, S$ are singlets.
The first term forces $Y \ne 0$ along the flat direction, and the
second term gives the singlet $S$ a mass as desired.

Another possibility for generating the $\mu$ term
is to add a singlet $S$ with superpotential couplings
at the weak scale
\beq
\label{singlet}
W = \lambda S H_u H_d + \frac{k}{3} S^3\ .
\eeq
If $S$ gets a VEV of order the weak scale, this can generate a $\mu$
term.
This model is appealing because it can relax the upper bound on the
lightest Higgs mass.

We can obtain a realistic version of this model in the context of the
`5--3--1' model presented above.
We assume that the
superpotential at the GUT scale (see Eq.~(\ref{superpot})) includes the
additional terms
\beq
\label{superpotadd}
\Delta W = \lambda S \bar{H} H + \frac{k}{3} S^3
+ \lambda' S \bar{T} T\ .
\eeq
It is easy to check that there is still a one-parameter flat direction
with $\avg{\Sigma}, \avg{\bar{\Sigma}}$ nonzero, and $\avg{S} = 0$,
in the approximation that SUSY is unbroken.
When SUSY is broken, the field $S$ will get a VEV at the weak scale
if $m_S^2 < 0$ at the weak scale.
($A$ terms will also tend to destabilize $S = 0$.)
The $S \bar{T} T$ coupling gives rise to a contribution
$\Delta m_S^2 \sim -\lambda'^2 g_3'^2$ at the
GUT scale, and so $m_S^2$ can have any value at the GUT scale.
Similar couplings can presumably be found in other GUT models.
It appears that there is no obstacle in constructing a realistic
model of this type.

Finally, one can generate a VEV for $S$ in~(\ref{singlet}),
and hence, a $\mu$ term, following the mechanism of ref.~\cite{kss}.
In this mechanism, $S$ is coupled to heavy fields of mass
$F_\phi/\lambda^\prime$, where $\lambda^\prime$ is a small number.
As was shown in~\cite{kss}, such a mass is easily generated
from AMSB, leading to a $\mu$ term and a $B$ term of the correct size.

\section{Phenomenology}\label{phenom}
We now turn to the superpartner mass spectrum.
Our analysis applies not just to the `5--3--1' model of Section~\ref{model},
but to general AMSB GUT models with the GUT scale determined by AMSB.

If the GUT gauge group is simple, the gaugino masses at the weak scale can
be derived by noting that $m_\lambda / g^2$ is RG invariant
to one loop, and using the GUT matching condition:
\beq\label{eq:trueGUTgaugino}
m_{\lambda i}(\mu)
= g_i^2(\mu) \frac{\beta_{\rm GUT}}{g_{\rm GUT}^3} F_\phi \ .
\eeq
Here $i = 1,2,3$ runs over the Standard Model group factors (with $SU(5)$
normalization for $g_1$) and
\beq
\beta_{\rm GUT} = \left.
\frac{d g_{\rm GUT}}{d \ln\mu} \right|_{\mu = \mgut}\ .
\eeq
In the `5--3--1' model the prediction for the gaugino masses is modified
by the extra group factors.
The matching conditions at the GUT scale are
\beq
\frac{m_{\lambda 3}}{g_3^2} =
\frac{m_{\lambda 5}}{g_5^2}
+ \frac{m_{\lambda 3'}}{g'^2_3}\ ,
\quad
m_{\lambda 2} = m_{\lambda 5} \ ,
\quad
\frac{m_{\lambda 1}}{g_1^2} =
\frac{m_{\lambda 5}}{g_5^2} +
\frac{m_{\lambda 1'}}{15 g'^2_1}\ ,
\eeq
where $g'_3$ and $g'_1$ are the $SU(3)$ and $U(1)$ couplings in the
`5--3--1' model.
We therefore recover the simple GUT prediction Eq.~(\ref{eq:trueGUTgaugino})
for the $SU(2)$ gaugino mass, while
\begin{eqnarray}
\label{eq:gaugino3}
m_{\lambda 3}(\mu) \!\!&=&\!\! g_3^2(\mu) \left[
\frac{\beta_5(\mgut)}{g_5^3(\mgut)} + \frac{\beta'_3(\mgut)}
{{g'_3}^2(\mgut)}
\right] F_\phi\ ,
\\
\label{eq:gaugino1}
m_{\lambda 1}(\mu) \!\!&=&\!\! g_1^2(\mu) \left[
\frac{\beta_5(\mgut)}{g_5^3(\mgut)} +
\frac{\beta'_1(\mgut)}{15 {g'_1}^2(\mgut)}
\right] F_\phi\ .
\end{eqnarray}
In the specific model presented in Section~\ref{model},
$\beta'_3 = 0$ (at one loop) and so the gluino mass is also given by the
simple GUT prediction.

We now consider scalars.
We first assume that $D$-term contributions to the scalar
masses are negligible.
In the `5--3--1' model, this is the case for $c-{\bar c}\ll 1$,
and $\lambda_2-\lambda_3\ll 1$ as discussed in Section~\ref{model}.
Scalar masses then unify into GUT
representations at the GUT scale.
This is of course familiar in traditional SUGRA models, but
does not happen
in minimal anomaly mediation, or in the modifications
discussed in Refs.~\cite{clmp,kss,rp}.

For the first two generations, and
for third-generation $\bar{\bf 5}$ fields for small $\tan\beta$,
one can neglect Yukawa couplings.%
\footnote{It is possible that these fields have large Yukawa couplings
with superheavy fields above the GUT scale.
This possibility is unmotivated, and is difficult to reconcile with the
requirement that the GUT group is asymptotically free.}
The AMSB soft mass at the GUT scale is then given by
the gauge contribution
\beq
\label{gutmass}
m^2_r(\mgut) = -\frac{C_{{\rm GUT},r}}{8\pi^2} g_{\rm GUT}
\beta_{\rm GUT} |F_\phi|^2\ ,
\eeq
where $C_{{\rm GUT},r}$ is the Casimir of the GUT representation of the field.
(Note that $m^2_r(\mgut) > 0$ if the GUT group is asymptotically free.)
The RG equations can then be used to compute the value at the weak scale.
The result is
\beq
m^2_r(\mu) = m^2_r(\mgut) + \sum_{i = 1}^3 \frac{2 C_{i,r}}{b_i}
\left( \frac{m_{\lambda i}}{g_i^2} \right)^2
\left[ g_{\rm GUT}^4 - g_i^4(\mu) \right]\ ,
\eeq
where
\beq
\frac{d g_i}{d\ln\mu} = \frac{b_i g_i^3}{16\pi^2}\ ,
\eeq
and $m_{\lambda i} /g_i^2$ is an RG invariant at one loop
(given by Eq.~(\ref{eq:trueGUTgaugino})
for a true GUT, or Eqs.~(\ref{eq:gaugino3}) and (\ref{eq:gaugino1})
in the `5--3--1' model).
Note that the running down to the weak scale gives a positive contribution
to the slepton masses, but the right-handed sleptons get a contribution
only from $U(1)_Y$, and this by itself is not large enough to allow
the slepton masses at the GUT scale to be significantly negative.
This motivates our choice of an asymtotically free GUT group.

For the third generation, the top Yukawa coupling affects the
predictions.
(For large $\tan\beta$ we must consider the bottom and tau Yukawa
couplings as well.)
The third-generation scalar masses at the GUT scale are given by
\beq
m_3^2 \sim +y_t^4 - y_t^2 g_{\rm GUT}^2 + y_t^2 \lambda^2\ ,
\eeq
where $\lambda$ is a cubic coupling involving the GUT Higgs fields.
Such couplings are expected generically to be present in order to
explain doublet-triplet splitting.
We see that there is a model-dependent positive contribution to the
third-generation scalar masses, so the only model-independent
prediction is a lower bound.
The value of $y_t$ is also senstive to $\tan\beta$.
Additionally, the $\mu$ term is important for the mass eigenstates of the stops
(and the sbottoms and staus for large $\tan\beta$) because AMSB gives large $A$
terms.
We have checked that for small $\tan\beta$, neglecting the $y_t^2 \lambda^2$
contribution at the GUT scale gives third-generation squark masses close
to that of the first two generations.
We conclude that there is no significant model-independent constraint on
the third generation scalar masses, other than the fact that scalar masses
of members of the same GUT multiplet unify at the GUT scale.

We now turn to the generic case, in which $D$-terms cannot be
neglected.
Different scalars then recieve model dependent contributions
to their masses-squared at the GUT scale.
For the `5--3--1' model, these contributions are proportional
to the hypercharge of the relevant scalar.
Eqn.~(\ref{gutmass}) then contains an additional term,
given by eqn.~(\ref{dtermcontribution}). In this case, we
cannot obtain an analytic expression for the soft masses
at low energies.
However, as explained in Section~\ref{model}, in a large
region of parameter space, the $D$-term contribution
does not modify the masses significantly, so that
the scalar masses-squared at low energies are still positive.

\section{Conclusions}\label{concl}
We studied grand unified theories (GUT's) in which supersymmetry breaking
in the visible sector is communicated by anomaly mediation.
We showed that if the GUT scale is determined by the vacuum expectation
value of a modulus field, then there is a simple and natural mechanism
that fixes the GUT scale at the value
\beq
M_{\rm GUT} \sim \frac{M_{\rm Planck}}{16\pi^2},
\eeq
independently of the couplings of the theory.
Because the GUT threshold is non-supersymmetric in such models,
superpartner masses are not on the anomaly-mediated
trajectory below the GUT scale.
This can be used to correct the main phenomenological problem
of `minimal' anomaly-mediated supersymmetry breaking, that is,
the tachyonic sleptons.
We showed that if the GUT group is asymptotically free
one can obtain a viable superpartner mass spectrum at low energies.
Also, the  resulting superpartner spectrum is very different from that
of previously studied AMSB models.
In particular, gaugino and scalar masses can unify, subject to corrections
from other group factors (for gaugino masses) and
$D$ terms (for scalar masses).

We demonstrated these ideas by constructing a GUT
model which includes a solution to the doublet-triplet
splitting problem.
The model we consider has some unappealing features, notably a Landau pole
not far from the GUT scale for a $U(1)$ factor.
However, we believe it is very interesting that
the physics of unification and supersymmetry breaking is so closely
intertwined in this class of models.
On the one hand, the GUT scale is determined dynamically
from the Planck scale by supersymmetry-breaking effects.
On the other hand, the GUT physics modifies superpartner
masses and leads to a viable low energy theory.
We believe that these ideas are worthy of further investigation.

\section{Acknowledgments}

Y.~Shirman thanks Takeo Moroi for useful discussions,
and the Aspen Center for Physics
for hospitality during the early stages of this project.
M.~Luty and Y. Shirman also thank the Santa Barbara ITP for hospitality during
the early stages of this project.
The work of M. Luty and E. Pont\'on was supported by
NSF grant \#PHY-98-02551.
The work of Z.~Chacko was supported by DOE contract DE-FG03-96-ER40956.
The work of Y. Shirman was supported by NSF grant
\#PHY-9802484. The work of Y.~Shadmi was supported by
DOE grants \#DF-FC02-94ER40818, and \#DE-FC02-91ER40671.

\nc{\ib}[3]{ {\em ibid. }{\bf #1} (19#2) #3}
\nc{\np}[3]{ {\em Nucl.\ Phys. }{\bf #1} (19#2) #3}
\nc{\pl}[3]{ {\em Phys.\ Lett. }{\bf #1} (19#2) #3}
\nc{\pr}[3]{ {\em Phys.\ Rev. }{\bf #1} (19#2) #3}
\nc{\prep}[3]{ {\em Phys.\ Rep. }{\bf #1} (19#2) #3}
\nc{\prl}[3]{ {\em Phys.\ Rev.\ Lett. }{\bf #1} (19#2) #3}

\end{document}